\documentclass{jps-cp}
\usepackage{txfonts} 

\title{Orbital dependent magnetic exchange interaction in CeX$_{\text c}$ (X$_{\text c}$=S, Se, Te)}

\author{Takeshi \textsc{Matsumura}$^{1}$ and Akira \textsc{Ochiai}$^{2}$}

\inst{$^{1}$Department of Quantum Matter, AdSM, Hiroshima University, Higashi-Hiroshima, Hiroshima 739-8530, Japan  \\
$^{2}$Department of Physics, Graduate School of Science, Tohoku University, Sendai 980-8578, Japan 
}

\email{tmatsu@hiroshima-u.ac.jp}

\recdate{September 15, 2019}

\abst{Dispersion relations of the crystal-field excitations in cubic antiferromagnets CeTe, CeSe, and CeS have been investigated by inelastic neutron scattering using single crystalline samples. 
The Fourier transform of the magnetic exchange interaction $J(\mib{q})$ obtained from the crystal-field dispersion is largely different from that of the mean-field interaction obtained from the N\'eel temperature and the Weiss temperature. 
From detailed reexamination of the magnetic susceptibility and these $J(\mib{q})$ relations, we conclude that the magnetic exchange interaction is dependent on the crystal-field levels. The interaction associated with the $\Gamma_8$ excited state is stronger than that with the $\Gamma_7$ ground state. 
}

\kword{Ce monochalcogenides, crystalline electric field, exchange interaction}

\begin{document}
\maketitle

\section{Introduction}
Ce monochalcogenides, CeX$_{\text c}$ (X$_{\text c}$ = S, Se, Te), crystallizing in the cubic NaCl-type structure, have been considered as simple antiferromagnets with the propagation vector $\mib{q}$=(1/2,1/2,1/2), where the magnetic degree of freedom arises from the $\Gamma_7$-doublet crystalline electric field (CEF) ground state \cite{Hulliger78,Ott79,Rossat-Mignod85,Donni93,Donni93b}. 
The N\'eel temperatures of 2.0 K, 5.4 K, and 8.4 K for CeTe, CeSe, and CeS, respectively, reflects the strength of the magnetic exchange interaction, which mainly depends on the inter-atomic distance between Ce atoms; the lattice constants are $a$=6.361 \AA\ for CeTe, 5.992 \AA\ for CeSe, and 5.776 \AA\ for CeS. 
The $\Gamma_8$-quartet excited state is well isolated at 32 K in CeTe, 116 K in CeSe, and 140 K in CeS, which are also associated with the lattice constant \cite{Rossat-Mignod85,Donni93b}.  
The ordered moments of 0.3 $\mu_{\text{B}}$ for CeTe, 0.56 $\mu_{\text{B}}$ for CeSe, and 0.57 $\mu_{\text{B}}$ for CeS are roughly consistent with 0.71 $\mu_{\text{B}}$ for the $\Gamma_7$ CEF eigenstate \cite{Ott79,Donni93b}. 
Although the reductions of the moments are considered to be due to the Kondo effect, which is the strongest in CeS and the weakest in CeTe, the reason for the most reduced value for CeTe has not been resolved yet \cite{Nakayama04a,Nakayama04b}. 

Recently, from the experimental studies of the physical properties of CeX$_{\text c}$'s extended to high pressures and high magnetic fields, the role of the $\Gamma_8$ excited state has been recognized \cite{Takaguchi15,Hayashi16}. 
In CeTe, with increasing pressure, a rich variety of magnetic phases appears in the $H-T$ phase diagram because of the fall in the $\Gamma_8$ energy-level under pressure, leading to more contribution of the magnetic and quadrupolar degrees of freedom of the $\Gamma_8$ quartet state. 
Even at ambient pressure, application of high magnetic field induces contribution from the $\Gamma_8$ excited state \cite{Nakamura19}. 
This is due to the relatively large off-diagonal matrix element between $\Gamma_7$ and $\Gamma_8$; 
$\langle \Gamma_7 | J_z | \Gamma_7 \rangle = \pm 0.833$, 
$\langle \Gamma_{8a} | J_z | \Gamma_{8a} \rangle = \pm 1.833$, $\langle \Gamma_{8b} | J_z | \Gamma_{8b} \rangle = \pm 0.5$, 
and $\langle \Gamma_7 | J_z | \Gamma_{8a} \rangle = \pm 1.491$. 
Also in CeSe and CeS, the $\Gamma_8$ energy-level falls with increasing pressure, which may be associated with the enhancement in the Kondo effect due to the increase in the orbital degeneracy.  
Although the $p$ -- $f$ mixing effect can be an important mechanism for the reduction in the CEF splitting, the detail has not been understood yet. 
It is therefore important to understand the nature of the magnetic exchange interaction in more detail to understand the anomalous properties observed under high pressures and high magnetic fields. 
For this purpose, in the present paper, we analyze the dispersion relations of the CEF excitation observed for CeX$_{\text c}$'s by inelastic neutron scattering experiments. 
The temperature dependences of the magnetic susceptibility are reexamined in detail to explain the observations consistently. 
It is shown that we need to introduce exchange interactions which is dependent on the CEF states.

\section{Experiment}
Single crystalline samples were prepared by the Bridgeman method using vacuum sealed tungsten crucibles \cite{Nakayama04a}. 
Inelastic neutron scattering experiments were performed using the triple-axis thermal neutron spectrometer TOPAN installed at the beam hole 6G of the research reactor JRR-3, Japan Atomic Energy Agency, Tokai. 
A monochromatized incident beam was obtained using the 002 Bragg reflection of PG crystals. 
The energy of the scattered beam was analyzed using a PG-002 analyzer in the constant final energy mode at 14.7 meV. 
Collimators were Blank-30'-30'-Blank. 

\section{Results and Analysis}
\begin{figure}
\begin{center}
\includegraphics[width=15cm]{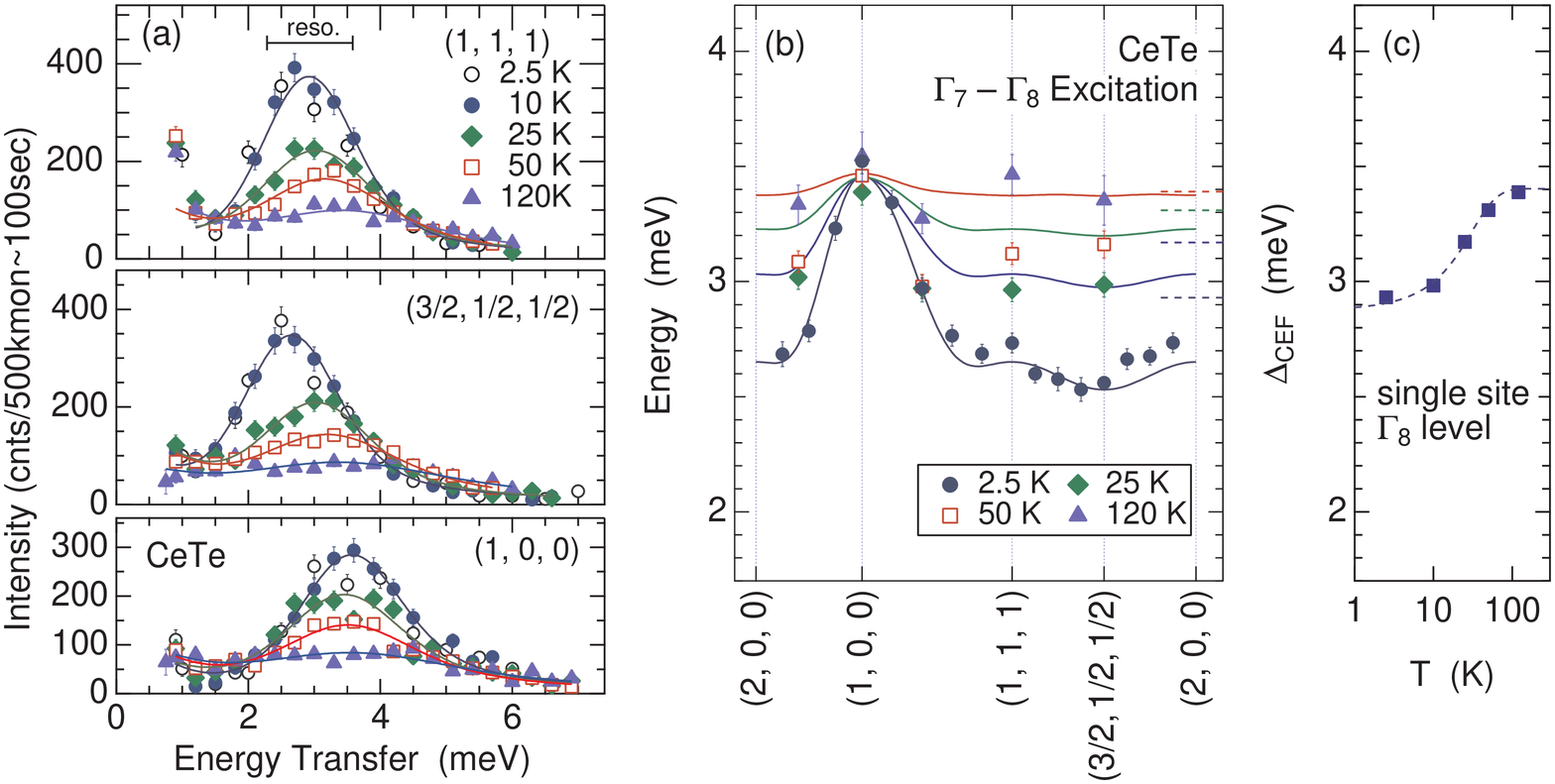}
\caption{(color online)  
(a) Inelastic neutron scattering spectra of CeTe measured at three reciprocal lattice points of $\Gamma$ : (1, 1, 1), L : (3/2, 1/2, 1/2), and X : (1, 0, 0). 
The instrumental energy resolution (full width at half maximum) at the energy transfer of 2.7 meV is represented by the horizontal bar in the top figure. The solid lines are the fits with a Lorentzian spectral function. 
(b) Dispersion relations of the $\Gamma_7$--$\Gamma_8$ CEF excitation in CeTe at several temperatures. 
The solid lines are the calculated dispersion relations by assuming a MF-RPA model with $J_1=0.26$ K and $J_2=-0.31$ K. 
The dashed lines represent the center of the dispersion, i.e., the single-site $\Gamma_8$ energy-level. 
(c) Temperature dependence of the single-site $\Gamma_8$ energy-level. 
}
\label{fig:CeTeINS}
\end{center}
\end{figure}

Figure \ref{fig:CeTeINS} shows the inelastic neutron scattering spectra of the $\Gamma_7$--$\Gamma_8$ CEF excitation of CeTe and the dispersion relations at several temperatures. 
The raw spectrum in Fig.~\ref{fig:CeTeINS}(a) was analyzed with a scattering function 
$S(E)=IP(E,\Delta,\Gamma)E/\{ 1-\exp (-E/k_{\text{B}}T) \}$, where $IP(E,\Delta,\Gamma)$ represents a Lorentzian spectral function centered at $E=\Delta$, the CEF excitation energy, with a half-width $\Gamma$ and an intensity $I$. The resultant parameters obtained from the fit is summarized in Fig.~\ref{fig:CeTeINS}(b). 
The $q$-dependence of the excitation energy is significant at low temperatures and becomes less dispersive at high temperatures. 
It is interesting that the energy is strongly dependent on temperature, whereas it is independent of temperature at the X-point (1,0,0) in the reciprocal space. 
The energy width of the excitation peak is almost resolution limited, indicating that the $4f$-electron is well localized in CeTe and the hybridization is weaker than those of CeSe and CeS. 

Let us analyze the dispersion relation by considering that it is caused by the inter-site magnetic exchange interaction $J_{ij}$.  
The neutron scattering function $S(\mib{q}, E)$ is related with the imaginary part of the generalized susceptibility $\chi(\mib{q}, E)$ by 
$S(\mib{q}, E) \propto \chi''(\mib{q}, E) / \{1-\exp (-E/k_{\text{B}}T) \}$, 
where $\chi(\mib{q}, E)$ can be assumed in the mean-field random-phase (MF-RPA) approximation by 
\begin{align}
\chi(\mib{q}, E) &= \frac{\displaystyle \chi_0(E)}{1-J(\mib{q}) \chi_0(E) /  g^2 \mu_{\text{B}}^{\;2} } \;. 
\label{eq:chiqE}
\end{align}
$J(\mib{q})$ represents the Fourier transform of the magnetic exchange interaction $J_{i,j}$ and 
$\chi_0(E)$ the frequency $(E=\hbar \omega)$ dependent single-site dynamic susceptibility due to the CEF split levels. 
By treating the $\Gamma_8$ energy level $\Delta$ and the exchange constants of $J_1$ (nearest neighbor) and $J_2$ (next nearest neighbor) as fitting parameters, we analyzed the dispersion relation of Fig.~\ref{fig:CeTeINS}. 
From the result at 2.5 K, $J_1=0.26$ K and $J_2=-0.31$ K were obtained. 
The temperature dependence of the $\Gamma_8$ energy level, which decreases with decreasing temperature, is shown in Fig.~\ref{fig:CeTeINS}(c). 
We do not understand the reason for this unusual behavior of $\Delta(T)$ at the moment.  

\begin{figure}
\begin{center}
\includegraphics[width=15cm]{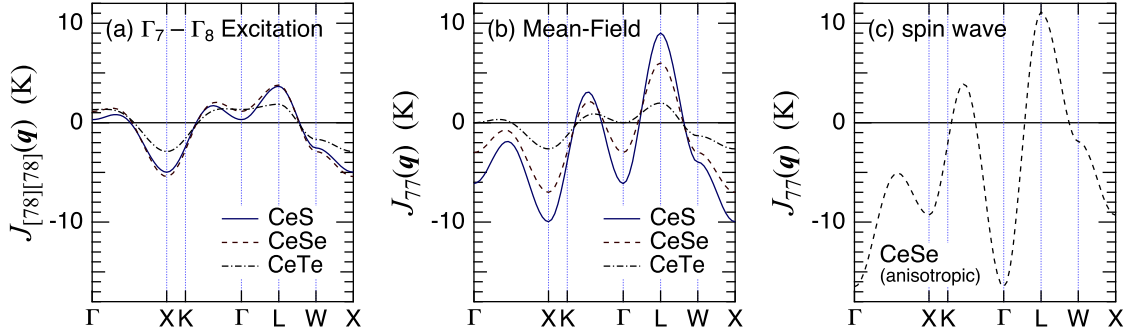}
\caption{(color online) 
(a) Dispersion relations of $J(\mib{q})$ obtained from the MF-RPA analysis of the CEF excitations for CeTe, CeSe, and CeS. 
(b) Dispersion relations of $J(\mib{q})$ to reproduce $T_{\text{N}}^{(\text{exp})}$ and $\lambda_{\mib{q}=0}^{(\text{exp})}$ by assuming $J(\mib{q}_{\text{L}})=-6J_2$ and $J(\mib{q}_{\Gamma})=12J_1 +6J_2$. 
(c) Dispersion relations of $J(\mib{q})$ deduced from the $J_1$ and $J_2$ parameters to explain the spin wave dispersion of CeSe in Ref.~\citen{Donni93}. 
}
\label{fig:CeXJQ}
\end{center}
\end{figure}

\begin{figure}
\begin{center}
\includegraphics[width=15cm]{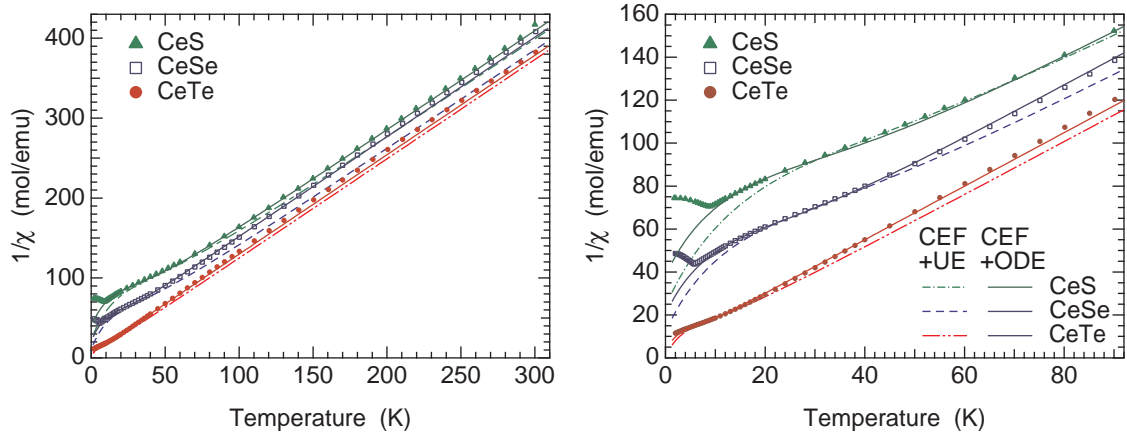}
\caption{(color online) 
Temperature dependences of inverse magnetic susceptibility of CeS, CeSe, and CeTe. The solid lines are the calculations considering the orbital dependent exchange (ODE) interactions for the CEF levels. The dashed, dot-dashed, and double dot-dashed lines are the calculations considering the  uniform exchange (UE) interaction to fit $1/\chi$ at low temperatures; $\lambda=-22$, $-12$, and $0$ (mol/emu) for CeS, CeSe, and CeTe, respectively. 
}
\label{fig:CeXMagsus}
\end{center}
\end{figure}

The dispersion relation of $J(\mib{q})=\sum_{j} J_j \exp (-i \mib{q}\cdot \mib{r}_j )$ deduced from the above obtained $J_1$ and $J_2$ values for CeTe is shown in Fig.~\ref{fig:CeXJQ}(a). The $J(\mib{q})$ relations for CeSe and CeS, which were also deduced from the same procedure as above by inelastic neutron scattering, are also shown in Fig.~\ref{fig:CeXJQ}(a). 
The exchange parameters are $(J_1, J_2)=(0.41, -0.63)$ K for CeSe and $(J_1, J_2)=(0.33, -0.61)$ K for CeS, which are summarized in Table I by $J^{\text{(CEF)}}$. 
For all the three compounds $J(\mib{q})$ takes the maximum at the L-point, which is consistent with the type-II magnetic order at (1/2,1/2,1/2). 
However, the calculated $T_{\text{N}}$'s from these $J(\mib{q}_{\text{L}})$ values, i.e., the temperature where $J(\mib{q}_{\text{L}})\chi_0(0)=g^2 \mu_{\text{B}}^{\;2}$ is satisfied in Eq. (\ref{eq:chiqE}), are different from the actual values. 
In addition, the positive $J(\mib{q}_{\Gamma})$'s in Fig.~\ref{fig:CeXJQ}(a), which represent the exchange interaction for a uniform field, is inconsistent with the apparently antiferromagnetic exchange observed in the inverse magnetic susceptibility shown in Fig.~\ref{fig:CeXMagsus}. 

Using the single-cite static susceptibility $\chi_0$ due to the CEF split levels, the static susceptibility $1/\chi(\mib{q})$ in the mean-field model is written as 
\begin{align}
\frac{1}{\chi(\mib{q})} = \frac{1}{\chi_0} - \frac{J(\mib{q})}{g^2 \mu_{\text{B}}^{\;2}} \;. \label{eq:chiqinv}
\end{align}
At the $\mib{q}$ vector where $J(\mib{q})$ takes the maximum $\chi(\mib{q})$ diverges at the highest temperature, 
which determines $T_{\text{N}}$ in the mean-field approximation. From the experimental $T_{\text{N}}$ and the relation of $J(\mib{q}_{\text{L}})=-6J_2$ for the fcc lattice, we can estimate the mean-field $J_2^{\text{(MF)}}$ value. 
The mean-field $J_1^{\text{(MF)}}$ is obtained from the shift of $1/\chi$ from $1/\chi_0$. 
Since $1/\chi(\mib{q}) = 1/\chi_0 - \lambda(\mib{q})$, where $\lambda(\mib{q}) = J(\mib{q})/g^2 \mu_{\text{B}}^{\;2}$, 
the mean-field $J(\mib{q}_{\Gamma})=12J_1 +6J_2$ is associated with the experimental $\lambda(\mib{q}$=$0)$. 
The resultant parameters are $(J_1^{\text{(MF)}}, J_2^{\text{(MF)}})=(0.165, -0.33)$ K for CeTe, $(0.24, -1.5)$ K for CeSe, and $(0.25, -1.0)$ K for CeS. 
The mean-field $J(\mib{q})$ curves for the three compounds calculated from these parameters are shown in Fig.~\ref{fig:CeXJQ}(b). 
$J(\mib{q}_{\text{L}})$ and $J(\mib{q}_{\Gamma})$ values in this figure reproduces $T_{\text{N}}$ and $\lambda$ just above $T_{\text{N}}$, respectively.

However, again, Eq.~(\ref{eq:chiqinv}), with a single parameter $\lambda$ assuming a uniform exchange constant, cannot well explain the temperature dependences of $1/\chi$. 
As shown in Fig.~\ref{fig:CeXMagsus} by the lines of uniform exchange, a simple shift of $1/\chi_0$ to fit the data at low temperatures fails to explain the data at high temperatures. 
The disagreement between $J^{(\text{CEF})}$ and $J^{(\text{MF})}$ also remains unresolved. 
To consider these problems, we take into account orbital dependent exchange interactions \cite{Aoki80}. 

The total magnetic moment $\langle \mu \rangle$ consists of the moments from the $\Gamma_7$, $\Gamma_8$, and from the Van-Vleck contribution through the off-diagonal matrix element, which are represented by $\langle \mu_7 \rangle$, $\langle \mu_8 \rangle$, and $\langle \mu_{[78]} \rangle$, respectively. 
The molecular field for $\langle \mu_7 \rangle$ arises from $\langle \mu_7 \rangle$ itself,  $\langle \mu_8 \rangle$, and $\langle \mu_{[78]} \rangle$, each of which has its own exchange constant, $\lambda_{77}$, $\lambda_{78}$, and $\lambda_{7[78]}$, respectively. The numbers without and with the brackets represent the Curie and the Van-Vleck terms, respectively. 
The molecular fields for $\langle \mu_8 \rangle$ and $\langle \mu_{[78]} \rangle$ are expressed in the same manner, where $\lambda_{87}$=$\lambda_{78}$, $\lambda_{7[78]}$=$\lambda_{[78]7}$, and $\lambda_{8[78]}$=$\lambda_{[78]8}$. 
Using the single ion susceptibilities $\chi_{0}^{(7)}$, $\chi_{0}^{(8)}$, and $\chi_{0}^{[78]}$ for the local CEF levels, the total magnetic susceptibility $\chi$ is obtained by solving the following equations:
\begin{align}
\langle \mu \rangle &= \langle \mu_7 + \mu_8 + \mu_{[78]} \rangle = \chi H \\
\langle \mu_7 \rangle &= \chi_{0}^{(7)} (H + \lambda_{77}\langle \mu_7 \rangle + \lambda_{78}\langle \mu_8 \rangle + \lambda_{7[78]}\langle \mu_{[78]} \rangle \\
\langle \mu_8 \rangle &= \chi_{0}^{(8)} (H + \lambda_{87}\langle \mu_7 \rangle + \lambda_{88}\langle \mu_8 \rangle + \lambda_{8[78]}\langle \mu_{[78]} \rangle \\
\langle \mu_{[78]} \rangle &= \chi_{0}^{[78]} (H + \lambda_{[78]7}\langle \mu_7 \rangle + \lambda_{[78]8}\langle \mu_8 \rangle + \lambda_{[78][78]}\langle \mu_{[78]} \rangle 
\end{align}

The parameters to fit the experimental $1/\chi$ in the whole temperature range from $T_{\text{N}}$ to 300 K are listed in Table \ref{tbl:1}. 
To avoid complexity, we assumed $\lambda_{77}=\lambda_{[78]7}$, $\lambda_{88}=\lambda_{[78]8}$, and $\lambda_{78}=(\lambda_{77}+\lambda_{88})/2$. 
As shown by the solid lines in Fig.~\ref{fig:CeXMagsus}, the data are well reproduced by considering the orbital dependent exchange constants. 
It is noted that the exchange constant for the CEF excitation correspond to $\lambda_{[78][78]}$ because the CEF excitation is associated with the off-diagonal elements between $\Gamma_7$ and $\Gamma_8$. 
The propagation of the CEF excitation (magnetic exciton) is caused by the exchange interaction between the Van-Vleck magnetic moments. 
Therefore, $J(\mib{q})$ in Fig.~\ref{fig:CeXJQ}(a) should be written as $J_{[78][78]}(\mib{q})$. 
The $\lambda_{[78][78]}$ parameters in Table \ref{tbl:1} are chosen so that they are consistent with the $J_{[78][78]}(\mib{q}_{\Gamma})$ values in Fig.~\ref{fig:CeXJQ}(a). 
On the other hand, $J(\mib{q})$ in Fig.~\ref{fig:CeXJQ}(b) should be written as $J_{77}(\mib{q})$ because the exchange interaction at low temperatures just above $T_{\text{N}}$ mostly arises between the $\Gamma_7$ ground states. 
By treating the exchange interactions as dependent on the CEF states, the difficulties in understanding the exchange constants and the magnetic susceptibilities were removed. 

\begin{table}
\caption{
The experimental results of $T_{\text{N}}$ (K) and $\lambda_{\mib{q}=0}$ (mol/emu), 
the nearest and next nearest exchange parameters $J_1^{(\text{CEF})}$ (K) and $J_2^{(\text{CEF})}$ (K) obtained from the fit of the energy dispersion of the CEF excitation by a MF-RPA model, $J_1^{(\text{MF})}$ (K) and $J_2^{(\text{MF})}$ (K) to explain  $T_{\text{N}}^{\text{(exp)}}$ and $\lambda_{\mib{q}=0}^{\text{(exp)}}$, $J_1^{(\text{SW})}$ (K) and $J_2^{(\text{SW})}$ (K) to explain the spin wave dispersion in the ordered phase, and the orbital dependent exchange parameters $\lambda_{77}=\lambda_{[78]7}$, $\lambda_{88}=\lambda_{[78]8}$, and $\lambda_{[78][78]}$ (mol/emu). 
}
\label{tbl:1}
\begin{center}
\begin{tabular}{cccccccccccc}
\hline
  & $T_{\text{N}}^{\text{(exp)}}$  & $\lambda_{\mib{q}=0}^{\text{(exp)}}$ & $J_1^{(\text{CEF})}$  &  $J_2^{(\text{CEF})}$  & $J_1^{(\text{MF})}$  &  $J_2^{(\text{MF})}$ & $J_1^{(\text{SW})}$  &  $J_2^{(\text{SW})}$ & 
  $\lambda_{77}$ &  $\lambda_{88}$  & $\lambda_{[78][78]}$   \\
\hline
CeS   &  8.4 & $-22$ & 0.33 & $-0.61$ & 0.24   & $-1.5$   &  --  &  --  & $-38$    & $-50$ & $1$ \\
CeSe & 5.4 & $-10$  & 0.41 & $-0.63$ & 0.25   & $-1.0$   &  $-0.46$  &  $-1.8$  & $-20$    & $-44$ & $4$ \\
CeTe  & 2.0 &     0     & 0.26 & $-0.31$ & 0.165 & $-0.33$ &  --  &   --  & $-3.0$   & $-10$ & $4$ \\
\hline
\end{tabular}
\end{center}
\end{table}

\section{Discussion}
The orbital dependent exchange is considered to play an important role in $f$-electron systems with crystal-field split levels. 
For example, in the filled-skutterudite compound SmRu$_4$P$_{12}$, only the $\Gamma_7$ state has a mixing with the conduction band of the $p$-electrons, which leads to a characteristic exchange interaction and the appearance of a magnetic-field induced charge-ordered phase \cite{Shiina13}.  
In the present analysis on CeX$_{\text{c}}$, it is remarked that the exchange interaction associated with the $\Gamma_8$ excited state is larger than that with the $\Gamma_7$ ground state. 
Although this may be associated with the fact that only the $\Gamma_8$ state has a mixing with the $p$-orbital of X$_{\text{c}}$, the details have not been clarified yet.  
Another point to be noted is that $J_2$ is larger than $J_1$, which can be concluded from Table. I regardless of the evaluation method. This also shows that the exchange interaction through the $p$-orbital of X$_{\text{c}}$ is more important than the nearest neighbor interaction through the conduction electron states of the Ce-$5d$ orbitals. 
The different exchange parameters obtained from the spin wave dispersion in CeSe is considered to be due to the change in the exchange interaction by forming the magnetic ordered state. 

The main contribution from the $J_2$ term is consistent with the formation of the type-II magnetic order with $\mib{q}$=(1/2,1/2,1/2), where all the moments at the second nearest neighbor sites are antiferromagnetically coupled. 
However, the moments on the nearest neighbor sites are frustrated in the fcc lattice. 
This could be one of the reasons for the reduced ordered moment in CeTe, where the $J_1$ and $J_2$ values are closer to the $J_2=-J_1$ line, the boundary between the type-II antiferromagnetic order and the ferromagnetic order in the mean-field model. 
This is a subject to be studied in future to clarify the effect of frustration.

\section{Summary}
We performed inelastic neutron scattering experiments to study the dispersion relations of the $\Gamma_7$-$\Gamma_8$ CEF excitations of CeTe, CeSe, and CeS using single crystalline samples. 
The results were analyzed using a mean-field random phase model and we obtained the exchange constants of $J_1$ and $J_2$. 
We analyzed the discrepancy between these values and those from the bulk properties of $T_{\text{N}}$ and $\lambda$ by introducing an orbital dependent exchange parameters, which succeeded in well explaining the temperature dependence of the inverse magnetic susceptibility. 
From these analyses, we suggest that the interaction associated with the $\Gamma_8$ excited state is stronger than that with the $\Gamma_7$ ground state, playing an important role in realizing the rich variety of ordered phases under pressure and high magnetic fields through the mixing with the $\Gamma_8$ excited state.

\section*{acknowledgement}
The authors are grateful to R. Shiina for useful discussions. 
This work was supported by JSPS KAKENHI Grant number 18K187370A. 
Magnetic susceptibility measurement using the MPMS were performed at N-BARD, Hiroshima University.


\begin{thebibliography}{9}
\bibitem{Hulliger78} F. Hulliger, B. Natterer and H. R. Ott, J. Magn. Magn. Mater. \textbf{8}, 87 (1978).
\bibitem{Ott79} H. R. Ott, J. K. Kjems, and F. Hulliger, Phys. Rev. Lett. \textbf{42}, 1378 (1979). 
\bibitem{Rossat-Mignod85} J. Rossat-Mignod, J. M. Effantin, P. Burlet, T. Chattopadhyay, L. P. Regnault, H. Bartholin, C. Vettier, O. Vogt, D. Ravot and J. C. Achart, J. Magn. Magn. Mater. \textbf{52}, 111 (1985). 
\bibitem{Donni93} A. D\"onni, A. Furrer, P. Fisher, S. M. Hayden, F. Hulliger and T. Suzuki, J. Phys: Condens. Matter \textbf{5}, 1119 (1993).
\bibitem{Donni93b} A. D\"onni, A. Furrer, P. Fisher, and F. Hulliger, Physica B \textbf{186-188}, 541 (1993). 
\bibitem{Nakayama04a} M. Nakayama, H. Aoki, A. Ochiai, T. Ito, H. Kumigashira, T. Takahashi and H. Harima, Phys. Rev. B \textbf{69}, 155116 (2004).
\bibitem{Nakayama04b} M. Nakayama, N. Kimura, H. Aoki, A. Ochiai, C. Terakura, T. Terashima and S. Uji, Phys. Rev. B \textbf{70}, 054421 (2004).
\bibitem{Takaguchi15} H. Takaguchi, Y. Hayashi, T. Matsumura, K. Umeo, M. Sera and A. Ochiai, J. Phys. Soc. Jpn. \textbf{84}, 044708 (2015).
\bibitem{Hayashi16} Y. Hayashi, S. Takai, T. Matsumura, H. Tanida, M. Sera, K. Matsubayashi, Y. Uwatoko and A. Ochiai, J. Phys. Soc. Jpn. \textbf{85}, 034704 (2016).
\bibitem{Nakamura19} S. Nakamura, S. Awaji, T. Yanagisawa, T. Saito, T. Matsumura, and A. Ochiai: this conference. 
\bibitem{Aoki80} Y. Aoki and T. Kasuya, Solid State Commun. \textbf{36}, 317 (1980).
\bibitem{Shiina13} R. Shiina, J. Phys. Soc. Jpn. \textbf{82}, 083713 (2013).
\end{thebibliography}
\end{document}